\documentclass{article} % use this for production
\usepackage[a4paper]{geometry}

\usepackage{amsfonts,amsmath,amssymb,amsthm}
\usepackage{booktabs}
\usepackage{multirow}
\usepackage{lipsum}
\usepackage{natbib}
\usepackage{graphicx}
\usepackage{hyperref}

\title{Spatial-Temporal Extreme Modeling for Point-to-Area Random Effects (PARE)}

\author{Carlynn Fagnant$^1$ \and Julia C. Schedler$^1$
\and Katherine B. Ensor$^1$\thanks{Corresponding author. Email: ensor@rice.edu}}
\date{$^1$Department of Statistics, Rice University, 6100 Main St., Houston, TX, U.S.A.}

\begin{document}

\maketitle

\begin{abstract}
One measurement modality for rainfall is a fixed location rain gauge. However, extreme rainfall, flooding, and other climate extremes often occur at larger spatial scales and affect more than one location in a community. For example, in 2017 Hurricane Harvey impacted all of Houston and the surrounding region causing widespread flooding. Flood risk modeling requires understanding of rainfall for hydrologic regions, which may contain one or more rain gauges. Further, policy changes to address the risks and damages of natural hazards such as severe flooding are usually made at the community/neighborhood level or higher geo-spatial scale. Therefore, spatial-temporal methods which convert results from one spatial scale to another are especially useful in applications for evolving environmental extremes. We develop a point-to-area random effects (PARE) modeling strategy for understanding spatial-temporal extreme values at the areal level, when the core information are time series at point locations distributed over the region. 

\end{abstract}

\section{Introduction}\label{sec:intro}
Spatial data can be observed over different spatial extents, leading to different types of spatial data and different statistical models. One type is point referenced data, or observations of the feature of interest indexed by a coordinate location. For example, whether or not a household experienced adverse outcomes during a major storm indexed by the latitude and longitude coordinates corresponding to the household. Modeling of point referenced data incorporates spatial dependence amongst the observations. For example, the covariance between two observations can be defined as a function of the distance between those two observations, and predictions of observations at new locations accounts for this spatial dependence. The consequences of ignoring spatial dependence includes biased estimates of standard errors, which can lead to incorrect inference. In this paper, we investigate rainfall data collected from fixed-location rain gauges in the Greater Houston area, revealing discernible spatial dependencies.

Even when spatial dependence in point-level data is properly accounted for, understanding regional spatial phenomena remains valuable for informed decision-making. Areal data, observed spatially across regions, often involves aggregating point-level data. For instance, in the case of the Hurricane Harvey Registry, responses collected at the point level (addresses) regarding adverse outcomes from the storm can be aggregated to the areal level (neighborhoods) to offer a city-wide summary of the storm's impacts \citep{miranda_texas_2021}. From a modeling perspective, incorporating spatial dependence in areal data requires specifying an adjacency matrix, indicating which regions are considered neighbors and thus inducing dependence between them. Various neighbor structures have been explored \citep{getis_constructing_2004}, with a common approach being to define neighbors as regions sharing a boundary point with a given region. The adjacency matrix can be weighted to form a spatial weight matrix, allowing a region's neighbors to exert spatially informed influences on its estimate in the model. For example, if region A has two neighbors, regions B and C, the impact of B and C on A need not be identical. If region B is larger than region C, it can be given a higher weight in the spatial weight matrix, enhancing its influence \citep{getis_spatial_2009}. Similar to point-referenced data, neglecting positive spatial dependence in areal data results in underestimation of parameter uncertainty and hence flawed statistical inference. In this study, we focus on analyzing data at the hydrologic region level, as these regions are scientifically meaningful in the context of rainfall \citep{acechouston_recommendation_2019}.

For our purposes, we observe time series of rainfall levels at fixed location gauges, but produce rainfall estimates for flood modeling and management decisions that are made at well-defined hydrologic areal regions. Our fundamental question is, how can the underlying distribution of rainfall be aggregated to the hydrologic level? This is called the change-of-support problem in spatial statistics \citep{gelfand_change_2001}. 

Much of the change-of-support literature focuses on obtaining estimates of the mean or median, in other words measures of center or ``typical'' values. In the context of flood risk, the interest is instead in estimating extreme events. When estimating risk, accounting for high-impact but lower probability tail events is of interest rather than the most likely scenario. Utilizing the hierarchical modeling framework advanced in \cite{craigmile_spatial_2014} and highlighted in \cite{cressie_statistics_2011}, we incorporate extreme value modeling, namely the generalized Pareto distribution and peak-over-threshold model (GPD+POT), as the data model to capture the extreme value nature of time series of point-level measured rainfall. We rely on asymptotic normality of the parameter estimates of the GPD for the Gaussian characterization of the process model in our hierarchy \citep{HoskingWallis1987}.

Our unique contribution is the synthesis of an end-to-end approach for estimating important regional hydrologic parameters in flood modeling, making use of ideas of hierarchical spatial modeling, the change-of-support framework as described in \cite{cressie_statistics_2011}, and conditional autoregressive (CAR) modeling to create a unique point-to-area-random-effects (PARE) model. Of particular importance is that our modeling is performed using spatial weights determined by the extended Hausdorff distance \citep{schedler_advances_2020}.

We apply three methods to move from point-referenced geostatistical data to achieve extremal inference at the areal level, namely our proposed PARE method, block kriging \citep{cressie_block_2006}, and a simple regional maximum approach. Block kriging is regularly used to address change-of-support issues in spatial modeling and the regional max model is often used by hydrologist studying extreme rainfall. We do not provide an extensive comparison of our PARE method with other strategies to address the spatial change-of-support issue. We apply our modeling paradigm to extreme rainfall for a large geographic region, specifically Houston, TX, with the goal of understanding the temporal evolution of extreme rainfall levels by hydrologic regions and accounting for spatial dependence. The methodology is adaptable to any problem where areal aggregation of point-level data is required.

\section{Understanding our Case Study, Hydrologic Regions, and Return Levels}

The findings of this paper hold significant implications for flood hazard mapping in the broader Houston, Texas area as well as subsequent analyses aimed at reducing flood risk. For example, they offer valuable insights into the fundamental design criteria for various hydraulic structures such as culverts, roadway drainage systems, bridges, and small dams. This study also complements recent literature that delves into attributing flood risk in Southeast Texas and coastal Louisiana to factors like climate change and urbanization \citep{Risser2017, vanderwiel2017, vanoldenborgh2017, wang2018, zhang2018, sebastian2019}, and it is particularly relevant given the recent release of NOAA Atlas 14 results for the State of Texas \citep{perica_precipitation-frequency_2018}.

\begin{figure}
    \centering
    \includegraphics[width=0.5\linewidth]{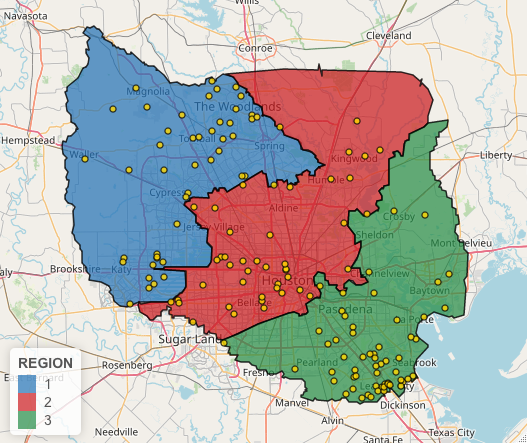}
    \caption{The three Harris County hydrologic regions are depicted by blue (Region 1), red (Region 2), and green (Region 3) areas. Rain guage locations for measured rainfall are represented by the yellow dots on the map.}
    \label{fig:Stations_map}
\end{figure}

To contribute to the scientific conversation in support of the resilience of the greater Houston, Texas, area we estimate rainfall return levels for three hydrologic regions depicted in Figure \ref{fig:Stations_map}. 

\subsection{Return Levels}
Return levels represent extreme rainfall statistics, serving to link a specific duration of rainfall with a defined time span. While the common duration of interest is daily or 24-hour rainfall, assessments of flood hazards can involve other durations such as 1-hour, 12-hour, 2-day, or 3-day intervals.

An $N$-year return level signifies the amount of rainfall expected to be equaled or surpassed, on average, once every $N$ years, where $N$ denotes the return period. This essentially indicates the rainfall magnitude with a probability of $\frac{1}{N}$, of being exceeded in any given year. 

It is important to recognize that the $N$-year return level is influenced by the duration of analysis. Thus, one would anticipate a 100-year 24-hour rainfall to exceed a 100-year 12-hour rainfall at a specific location. For our purposes, we consider 25-year, 100-year, and 500-year 24-hour rainfall return levels. These levels are often referred to as the 25-, 100-, and 500-year flood plane levels.

Our estimated return levels, with uncertainty bounds, are derived from our spatial hierarchical modeling, coupled with extreme value models as the data model for the time series of rainfall measurements at each rain gauge. These estimated return levels are key inputs for the large scale flood models in operation for the greater Houston area \citep{fagnant_characterizing_2020}.  

\subsection{Data Collection and Processing}
\label{sec:dataProcess}

We derive historical precipitation data from NOAA's Global Historical Climatology Network (GHCN)-Daily product version 3.22 \citep{Menne2018}, a comprehensive database aggregating daily observations from land surface stations worldwide. These data undergo stringent quality assurance procedures and are procured from various National Meteorological and Hydrological Centers (NMHCs) globally. Station records range from less than a year to over 175 years. Our analysis focuses on a study area delineated within coordinates approximately 28.1°N to 31.0°N, 92.7°W to 97.0°W, coinciding with specified hydrologic regions shown in Figure \ref{fig:Stations_map}. Within this area, we identified 149 stations providing data from January 1, 1900, to December 31, 2020. We compiled a dataset comprising daily precipitation totals from these 149 stations during this period, accessible at \href{doi.org/10.25612/837.XVGJ30NMA45X}{doi.org/10.25612/837.XVGJ30NMA45X} [\cite{UnitedStatesNationalOceanicAndAtmosphericAdministration2020SoutheastPrecipitation}].

Subsequent to initial data collection, the rainfall time series undergo processing to identify independent rainfall events. As the extreme value theory we employ assumes the independence of each occurrence, preprocessing is necessary to mitigate correlations between events. Such correlations arise from single storm events generating multi-day precipitation, wherein more than one day surpasses the specified threshold for our models. In such cases, despite the threshold being exceeded multiple times, there exists only one storm occurrence; hence, we retain only the highest 24 hour total from that storm. In simpler terms, if two or more consecutive days exhibit nonzero rainfall, only the day with the highest rainfall is retained, while the others are set to zero. Following this declustering process, the data are rendered independent, thus suitable for subsequent peaks-over-threshold analysis. This data adjustment, known as declustering, is a widely used technique in rainfall time series analyses \citep{Gilleland2016ExtRemesR, fagnant_characterizing_2020}. 

\section{Methods}\label{sec:methods}
The models presented are hierarchical, consisting of three steps: an extreme value modeling step, a spatial modeling step, and repeating the previous two steps for 3-different time epochs. In the data-process-parameter model approach advocated in \cite{cressie_statistics_2011}, the extreme value modeling serves as our data model, and then the spatial structure of step two is captured in our process model. Although the data will have a spatial index for all steps, spatial structure is only modeled in the second step. 
\subsection{Extreme value modeling}\label{EVM}
Our starting point is the time series observed at each of $n$ spatial locations. A first step is to characterize each of these time series in an extreme value modeling context appropriate for the application. For this paper, the application concerns rainfall in Southeast Texas, which has been successfully modeled using the generalized-Pareto peak-over-threshold distribution (GPD+POT) to a contiguous segment of each time series.  A primary reason for the choice of GPD+POT is the fact that the Houston region experiences multiple extreme rain events within a year. The GPD+POT applied to 24-hour rainfall events allows for this unique feature for the region. A detailed characterization of the choice of this particular extreme value model for the problem at hand is the focus of the paper \cite{fagnant_characterizing_2020}.

The peaks-over-threshold (POT) approach selects all events above a threshold for better return period estimation, incorporating a broader range of rainfall events compared to say the block maxima method, which is another extreme value modeling approach. POT treats independent rainfall occurrences as identically distributed random variables and focuses on observations exceeding a specified threshold $u$ to analyze extremes, resulting in a conditional distribution:

\begin{equation}
    P(X>u+y|X>u) = \frac{1-F(u+y)}{1-F(u)}, \quad  y>0
\end{equation}

The conditional distribution converges to the generalized Pareto distribution (GPD) of the form: 
\begin{equation}
    P(X>x|X>u) = \Bigg[1+\frac{\xi(x-u)}{\widetilde{\sigma}}\Bigg]^{-1/\xi}, \quad x>u
\end{equation}
where $\xi$ is the shape parameter. The scale parameter for the GPD+POT is a function of the generalized extreme value parameters and the threshold, namely $\big(\widetilde{\sigma} = \sigma + \xi(u-\mu)\big)$. Here $\mu$ is the location parameter and $\sigma$ is the scale parameter for the Generalized Extreme Value (GEV) distribution. These are asymptotic relationships that hold for appropriate thresholds. 

The GPD+POT approach, widely adopted in recent extreme rainfall analyses, owes its popularity to its capacity to encompass temporally precise rainfall data and its theoretical linkage to the GEV distribution. 

Return levels for each rain gauge time series can be calculated directly from the fitted distribution parameters ($\sigma, \xi$, and threshold $u$) obtained during the peaks-over-threshold modeling, along with the estimated rate of occurrences over the threshold. We utilize the R package {\tt extRemes} \citep{Gilleland2016ExtRemesR} to fit our GPD+POT and to calculate return levels and corresponding confidence intervals. A detailed characterization of the choice of GPD+POT, plus selection of the threshold to one-inch of rainfall is the provided in \cite{fagnant_characterizing_2020}. 

For our spatial model rainfall case study we focus on three time epochs each of forty years, namely 1920 through 1960, 1950 through 1990, and 1990 through 2020. For each temporal window and each location, we obtain estimates for the parameters of shape, scale and rate of the GPD+POT. These GPD+POT model estimates can be viewed in the hierarchical modeling paradigm advanced in \cite{cressie_statistics_2011} as well as \cite{cooley_survey_2012}, where we consider the data model, process model and predictive distribution. In this case, the GPD+POT serves as the data model for our rainfall time series. 

\subsection{Spatial Models}\label{}
For each of our three temporal windows we model the point-to-area spatial structure using three different spatial paradigms. These paradigms include our contribution of Point-to-area Random Effects (PARE) incorporating the extended Hausdorff distance, traditional Block Kriging, and the Regional Max model. The statistical literature on change of support is extensive. In addition, we compare our PARE model to classic approaches used by engineers working in flood modeling (see ACEC estimates described in \ref{sec:results}). Each paradigm is described below. 

\subsubsection{Model 1. Point-to-area Random Effects (PARE)}\label{sec:PARE}
For the first model, we make use of ideas behind the change of support framework as described in \cite{cressie_statistics_2011} and conditional autoregressive (CAR) modeling \cite{besag_spatial_1974}to create a unique random effects model to move from the point-level to the area-level. Of particular importance is that our modeling is performed using weights determined by the extended Hausdorff distance. 

A CAR model is a popular choice for accounting for spatial correlation when working with areal or lattice data, and can easily be extended to find the relationship between covariates measured at the same areal level. The CAR model accounts for spatial dependence between the areal units by specifying an $r \times r$ spatial weight matrix, where $r$ is the number of regions. If the $ij^{th}$ entry of the spatial weight matrix is nonzero, regions $i$ and $j$ are considered ``neighbors'', and the CAR model will use observations at region $i$ to inform estimates at region $j$ and vice-versa. Determining which regions are neighbors is often based on shared boundary points (contiguity) or centroid distance (e.g., k nearest neighbors or inverse distance weighting). Instead, we apply a method derived from the Hausdorff distance, a distance metric defined on sets. Specifically, we pursue a specification for the weight matrix which uses the inverse of the median Hausdorff distance between regions, yielding neighbor weights based on the maximum of the median distances between each pair of regions. The median Hausdorff distance accounts for irregularities in the geometry or orientation of the regions by only considering the closest 50\% of a region when computing pairwise distances. The median was chosen to give distances comparable to the commonly used centroid distance.  \citep{schedler_juliaschedlerhausdorff_2020}.

We make use of the covariance structure defined in a CAR model for areal data, but instead bring everything to the dimension of the stations (point-level observations). In order to create an $n \times n$ weight matrix, we must define distances from each point to the others. However, since we are interested in moving to the areal level, we instead define point-to-point distances by the associated area-to-area distance of the regions the points fall within. Particularly, we use the extended Hausdorff distance with $f = 0.5$ (in other words, the median Hausdorff distance) between our three regions to define these distances. For example, the distance between a point in region 1 to a point in region 2 is defined as the median Hausdorff distance between regions 1 and 2. This means that there will be many repeat values in our distance matrix for every pair of points that fall within the same pair of regions. The matrix will look like a block version of the region-to-region distance matrix where each block is made up of a single repeated value. 

Specifically, for the distance matrix, let $D$ be the $3 \times 3$ matrix defining the median Hausdorff distance $d$ between the three regions. $D$ can be found using the {\tt hausMat} function in the {\tt hausdorff} package \citep{schedler_juliaschedlerhausdorff_2020}. Let $n_j$ be the number of stations in region j. Then $D$ and $D_{block}$ are defined as follows,
\begin{equation}
D = 
\begin{bmatrix}
d[1,1] & d[1,2] & d[1,3]\\
d[1,2] & d[2,2] & d[2,3]\\
d[1,3] & d[2,3] & d[3,3]
\end{bmatrix}  
\end{equation}

\begin{equation}
D_{block} = 
\begin{bmatrix}
C_{n_1\times n_1} & D[1,2]_{n_1\times n_2} & D[1,3]_{n_1\times n_3}\\
D[1,2]_{n_2\times n_1} & C_{n_2\times n_2} & D[2,3]_{n_2\times n_3} \\
D[1,3]_{n_3\times n_1} & D[2,3]_{n_3\times n_2} & C_{n_3\times n_3}
\end{bmatrix}    
\end{equation}
where each block in $D_{block}$ is a constant value times a matrix of 1's of the specified dimension. For example, $D[1,2]_{n_1\times n_2} = d[1,2] * \mathbf{1}_{n_1\times n_2}$. The diagonal elements of $D_{block}$ are identically a constant $c$, which we now define.

While the distance from a region to itself is technically zero, we define the distance between points within the same region to be a specified positive constant $c$ that is less than the other region-to-region distances (i.e. $c < d[1,2], \,d[1,3]$). We do this because there is spatial variability of points within each region. This construction ensures the weight matrix is invertible and still weights values within a region more than values from other regions. For our application, we choose $c=1$ mile.\footnote{The choice of $c$ does not influence our final return level estimates. A sensitivity analysis is provided in \cite{fagnant_spatiotemporal_2021}.}  The distance matrix is converted into a weight matrix $W$ by taking the reciprocal of each element in the matrix in order to create inverse-distance weights. Overall, with the block structure of our weight matrix $W$, the observations within each region are weighted equally, but less than those stations within regions closer to the target region or within the target region itself. When using $c=1$, the inverse distance weight matrix has all elements $\leq 1$. If using a different value for $c$, one could scalar normalize the matrix by dividing all entries of the matrix by the value of the maximum entry.

We bring in the change-of-support concept in the mean structure of the model through covariates by creating variables which are indicator functions describing which region each station falls within. Let $\mathbf{Z}$ be the vector of point-level observations (extreme value parameter estimates for each rain guage location) at stations $\mathbf{s} = (s_1,\dots,s_n)$. 
Let $\mathbf{Y}$ be the process of interest measured at the spatial areas/regions $\mathbf{r} = (r_1, r_2, r_3)$. The $n\times3$ matrix $\mathbf{H}$ represents indicator variables for the regions the station falls within, in other words, 
\begin{equation}
 h_{ij} = \begin{cases}
1 & \text{if }\; s_i \in r_j \\
0 & \text{otherwise}.
\end{cases}   
\end{equation}

The traditional areal CAR model with covariates is defined through the following structure:
\begin{equation}
Y_j | y_{(-j)} \sim N\Big( x_j \beta + \underset{k\neq j}{\sum} \rho w_{jk} (y_k - x_k\beta), \;\; m_{jj}\Big)
\end{equation}
where $y_{(-j)}$ represents all areal values $y$ except for $y_j$, $\;\beta$ is the coefficient for covariate $x$,  $\;\rho$ is a spatial dependence parameter, $w_{jk}$ are entries from the weight matrix $W$, and $m_{jj} = \tau_j^2$, the conditional variance of region $j$. 

We will use a similar structure, but instead define it on the point-level. As such, we use it to model the observations $\mathbf{Z}$ instead of the area-level process values $\mathbf{Y}$. Our covariates $\mathbf{x}_i$ are represented by the matrix $\mathbf{H}$, so we can set $X=\mathbf{H}$. Our proposed point-to-area random effects (PARE) model takes the following structure:
\begin{equation}
Z_i | z_{(-i)} \sim N\Big( \mathbf{x}_i\:\!' \,\mathbf{\beta} + \underset{j\neq i}{\sum} \rho w_{ij} (z_j - \mathbf{x}_i\:\!' \,\mathbf{\beta}), \;\; m_{ii}\Big)
\end{equation}
which can also be written as $Z \sim N(X\mathbf{\beta}, \,\Sigma_{CAR})$ where $\Sigma_{CAR} = (I-\rho W)^{-1} M$ and $M = diag(\tau_1^2, \dots, \tau_n^2)$. This model can be run in R using the {\tt spautolm} function from the {\tt spatialreg} package \citep{bivand_applied_2013} with family = ``CAR''. The required inputs include observations $\mathbf{Z}$, covariates $X=\mathbf{H}$, and the weight matrix $W$. The model output provides maximum likelihood estimates for $\rho$ and $\mathbf{\beta}$. We interpret the coefficient estimates $\mathbf{\hat{\beta}}$ to be the estimates of our process $\mathbf{Y}$ at the areal level, i.e. the GPD scale, shape, and rate parameter estimates for the three regions.

One assumption the CAR model requires is that the covariance matrix $\Sigma_{CAR}$ be symmetric. Additionally, since $\Sigma_{CAR}$ is defined as $(I-\rho W)^{-1} M$, our weight matrix $W$ must be invertible. With the current block setup for the weight matrix, our $W$ is indeed symmetric. However, by construction our $W$ has many rows and columns with the same exact values, thereby making the matrix singular and non-invertible. We address this issue through jittering, in other words, adding a normal random variable with small standard deviation of 0.1 to the extended Hausdorff distance matrix which has values 7.7, 7.9, and 27.6 for each of the three pairwise distances. The jittered matrix is made symmetric by replacing the upper trianglur values with the lower triangular values, or vice versa. 

\subsubsection{Model 2. Block Kriging}\label{sec:bkrige}

For the second model, ordinary kriging is performed on the point-level extreme value estimates to obtain estimates of the extreme value parameters on a uniform fine grid. Integrating over the gridded points within each region reveals the ``block'' average, or the overall parameter estimate for each region. This approach is an established technique in the change-of-support literature to address the transition of spatial data from point-to-area \citep{gotway_combining_2002,craigmile_spatial_2014}, and is known more generally as block kriging \citep{cressie_block_2006}. We present this more established approach from the spatial literature so that we may compare it against our proposed PARE model.

We start by defining this block kriging model in terms of an underlying hierarchical model framework, and later describe how we estimate it in practice.
We follow the notation of \cite{cressie_statistics_2011}, but present it in terms of our application.

We summarize the hierarchical framework behind kriging of our GPD+POT parameter estimates below, where the predictive distribution gives the kriging estimates for new location $s_0$.
In particular, $Y^*(s_0)$ is the ordinary kriging predictor and $\sigma_Y^2(s_0)$ the associated kriging variance, which are given below in equations (\ref{eqn:krigeVar}). 

In this paradigm, we let $\mathbf{Z}$ be the vector of point-level observations (extreme value parameter estimates) at stations $\mathbf{s} = (s_1,\dots,s_n)$.
Let $\mathbf{Y}$ be the process of interest assessed on a uniform grid of points $G$ spanning the regions. The data process $Z$ is considered a noisy version of the underlying process $Y$ with measurement error, i.e. $Z = Y + \epsilon$ where $\epsilon \sim N(0, \sigma_\epsilon^2)$.
Therefore we have the data model as $[Z(s_i) | Y(s_i), \sigma_\epsilon^2] \sim N\left( Y(s_i), \sigma_\epsilon^2 \right)$.

Next we model the process at point $s$, or $Y(s)$ with constant but unknown mean $\mu_Y(s) = \mu$ and covariance $C_Y(u, v) = {\rm Cov}\left( Y(u), Y(v) \right)$ for $s, u, v \in D_s$. Then the process model is: $[Y(\cdot) | \mu, C_Y]\sim N\left( \mu, \; C_Y \right)$.

Finally, the predictive distribution for a new point $s_0$ is given by:
\begin{equation}
\begin{split}
    Y(s_0) | \mathbf{Z}, \mu, C_Y, \sigma_\epsilon^2] &\propto [Z(s_i) | Y(s_i), \sigma_\epsilon^2] [Y(\cdot) | \mu, C_Y] \\
    &\sim N\left( E[Y(s_0) | \mathbf{Z}], var(Y(s_0) | \mathbf{Z})\right)\\
&\sim N\left(Y^*(s_0), \sigma_Y^2(s_0)\right)
\end{split}
\end{equation}
where
\begin{equation}
\begin{aligned}
Y^*(s_0) &= \mu + \mathbf{c}_Y(s_0)'C_Z^{-1}(\mathbf{Z}-\mathbf{\mu}\mathbf{1}),\\%\label{eqn:krigeMean}
\sigma_Y^2(s_0)& = C_Y(s_0, s_0) - \mathbf{c}_Y(s_0)'C_Z^{-1}\mathbf{c}_Y(s_0)  + \\
&\quad \left(1 - \mathbf{1}'C_Z^{-1}\mathbf{c}_Y(s_0) \right)^2 / \left(\mathbf{1}'C_Z^{-1}\mathbf{1} \right)
\label{eqn:krigeVar}
\end{aligned}
\end{equation}
Defining the mean and covariance elements for $Y$ we have:
\begin{equation}
    \begin{aligned}
        C_Y(u,v) &= cov(Y(u), Y(v)) \text{ for } u, v \in D_s,\\
\mathbf{c}_Y(s_0)' &= cov(Y(s_0), \mathbf{Z}) = \left( C_Y(s_0, s_1), \dots, C_Y(s_0, s_n) \right)\\
&\text{and}\\
\mu = \hat{\mu}_{gls} &= \left(\mathbf{1}'C_Z^{-1}\mathbf{Z} \right) / \left(\mathbf{1}'C_Z^{-1}\mathbf{1} \right).    
\end{aligned}
\end{equation}

Finally, $C_Z$  is an $n\times n$ matrix where 
\begin{equation}
 C_Z(u,v) = \begin{cases}
C_Y(v,v) + \sigma_\epsilon^2 & \text{if }\; u=v\\
C_Y(u,v) & \text{if }\; u\neq v
\end{cases}\\   
\end{equation}

These equations for the ordinary kriging predictor follow the notation of \cite{cressie_statistics_2011}. Note that the constant mean $\mu$ is taken to be the generalized least squares estimate, $\hat{\mu}_{gls}$.

The next step in block kriging is to predict at a uniform grid $G$ across the three hydrologic regions, giving a (discretized) predicted spatial surface of the entire region. We then average the predicted values over the respective regions  $\hat{y}(r_j) = \frac{1}{n_{r_j}} \underset{g_k \in G\cap r_j}{\sum} y(g_k)$ where $n_{r_j}$ is the number of grid points in region $j$.

Turning to our specific application, we are interested in bringing extreme value analysis from the point observation level to the level of our hydrologic regions. Similar to the PARE model, we do this by bringing the extreme value parameter estimates to the region level and then calculate return levels based on the region level parameter estimates. Therefore, the kriging model above is performed where the observations are taken to be the extreme value parameter estimates for each station based on the univariate extreme value modeling described in \ref{EVM}.

One added benefit to the block kriging model not yet present in the PARE model is that kriging can be performed on multivariate data, meaning that the parameters can be modeled jointly if appropriate. In exploratory data analysis we find that the shape and log(scale) parameters are negatively correlated with each other, while the rate parameter does not show a relationship to either. In order to capture these relationships, we estimate the model above by cokriging the shape and log(scale) parameters jointly and kriging the rate parameter separately. The {\tt gstat} package in R provides functions to fit a cross-variogram and perform cokriging \citep{pebesma_multivariable_2004}.

\subsubsection{Model 3. Regional Max}\label{sec:regmax}
We discuss one final model, called the regional max model, which is a more simplified approach towards obtaining return level estimates for each region. This model is a data analytics approach one might take in order to summarize extreme data for a region. In particular, it takes the maximum daily rainfall value across all stations within a region to create a consolidated series of these maximum daily rainfall totals for each region, hence the name regional max. From the consolidated series, we can then perform traditional univariate extreme value analysis using the GPD+POT to obtain return levels estimates for each region.

We note that this model is a data analytic approach for which the mathematical theory has not been developed. Additionally, the Regional Max model is only spatial in the sense that maximums are taken over spatial subsets (the regions)-- the spatial structure is not modeled directly. We include it as a simple comparison against our proposed PARE model, as well as the traditional Block Kriging approach to change-of-support. We predict that since the regional max model uses the daily maximum values in combination with the GPD+POT, it will likely produce higher return level estimates than the previous models. 

\section{Simulations}

Recall that the purpose of the models above is to bring information and extreme value modeling estimates from the point-referenced observations to the areal level of the three specific hydrologic regions in the greater Houston area. In order to test the performance of the models proposed, we simulate multiple data sets of similar spatial structure to run the models on. We simulate data assuming we know the true extreme value parameters at the hydrologic region level. This simulation paradigm is consistent with the requirements from the hydrologists, in that the goal is to understand the 25-, 100-, and 500-year return levels for each of the three regions. From these simulated data sets, in which we set the true values for each hydrologic region, we can test how well each spatial model paradigm recovers these true values.

In order to construct simulated data, we start from the end with our estimated true values, and work backwards step-by-step to create data that results in the proper structure. The first step taken is to observe the mean structure of the extreme value parameters of the data that we already have. Taking the GPD fits of the last 40 years of data (1981-2020) for the stations within the three hydrologic regions, we find the mean values of the shape and scale parameter estimates by region. We set these as the true parameter values for the regions, and aim to recover similar values when we run our models on the simulated data. 

\begin{figure}[!ht]
    \centering
    \includegraphics[width=3in]{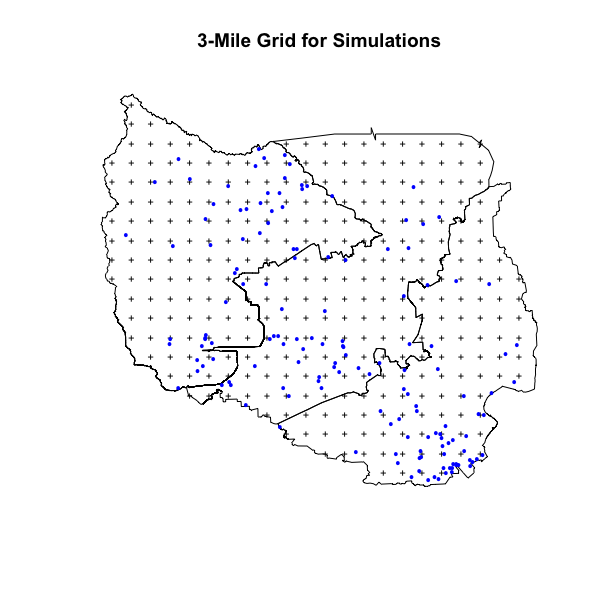}
    \caption{Map of the three hydrologic regions in Harris County (outlined in black), the station locations with data available for the 1981-2020 period (blue dots), and the grid locations where the simulated rainfall data are generated to (+ symbols).}
  \label{fig:sim_grid}
\end{figure}

In particular, the simulated rainfall data are generated on a uniform grid  with a resolution of three miles. This produces 314 points of daily rainfall data within the three hydrologic regions. The simulation grid is visible in Figure \ref{fig:sim_grid}. All points falling within a region are assigned the same true shape and scale parameters, which are the average values calculated from the last 40 years of observed data. 

The rate parameter is held constant in our simulation, which is supported by the result of fitting our three models to the rainfall data in Section \ref{sec:results}. In particular, the rate parameter has the least deviation between regions or models, staying fairly constant across the larger area. The constant rate is set to be 0.0544, the observed average of the rate parameter across all stations in the last 40-year period (1981-2020).

For the next step, we use the parameter estimates at each location to simulate daily rainfall data from the GPD. Since we use the GPD to model independent exceedances above the threshold, we can similarly only generate data values that are independent observations above the threshold. As such, we only generate a proportion of the daily data-- a proportion which is equal to our constant rate. For a 40-year period, we generate 795 daily values. In order to preserve the same rate parameter estimates for our fit, we fill the remaining daily values with zero representing a rain value below the threshold. %For the remaining data to fill... 
We generate rainfall values (exceeding the threshold) from the GPD using the {\tt revd} function from the {\tt extRemes} package in R \citep{Gilleland2016ExtRemesR}. % add equation?

The generated rainfall data lacks time-reference since we did not assign the data values to specific dates. We can only guarantee that the generated observations are independent. The lack of time-referencing only poses a problem for the regional max model, since it works off of the raw time-referenced daily data and not the GPD fits at the station level. In the interest of evaluating the performance of the regional max model in capturing the true parameter values, we propose an additional step to reorder the simulated rainfall values to impose a pseudo-time-referencing. This approach makes use of the idea that for one day, if there is a large rainfall amount at one station, the other stations are more likely to also have larger rainfall amounts that day. In general, rainfall totals across stations in a single day will be correlated, though not perfectly. To impose temporal considerations, we rank each simulated series and then perturb the ranks with uniform random noise, and then re-rank each series to assign a temporal structure. This strategy induces temporal correlation in the simulated series. Therefore, the regional max approach can be implemented, taking the maximum across each "day" based on the region each station falls within.

\subsection{Results from Simulation Study}\label{sec:sim_results}

Fifty iterations of the simulation are performed and the results displayed in Table \ref{tab:simTable}. Model performance is evaluated using two metrics: the Root Mean Square Error (RMSE), and the Mean Absolute Error (MAE). If $\hat{\xi}_1, \hat{\xi_2}, \dots, \hat{\xi}_{50}$ are the 50 estimates of the shape parameter for a given model, $\xi$ is the true value of the shape parameter for the simulation, then the RMSE and MAE are calculated as follows:

\begin{equation}
\begin{aligned}
    RMSE &= \sqrt{\frac{\sum_{i = 1}^{50}(\hat{\xi}_i - \xi)^2}{50}} \\
    MAE &= \frac{\sum_{i = 1}^{50}\vert \hat{\xi}_i - \xi \vert}{50}.
\end{aligned}
\end{equation}

We obtain values of both the RMSE and MAE the the shape and scale parameters for each combination of the three regions and each of the three models. 

While the first two models both performed well at recovering the true parameter values, our PARE Model performs almost uniformly better than the block kriging model according to the RMSE and MAE. The block kriging model only performed better in estimating the scale parameter for region 3, according to the RMSE. The regional max model performed much worse at capturing the true parameter values, especially that of the scale, just as we expected after seeing the large scale estimates in the results on observed data (see Table \ref{tab:paramByRegion}). The RMSE and MAE values for the regional max were around 10--30 times larger than those of the other two models.

\section{Case Study Results}\label{sec:results}
We apply the proposed models to our rainfall data for the last 40 years (1981-2020) and compare the regional extreme value estimates produced by each. The comparison of parameter estimates and return levels are available in Tables \ref{tab:paramByRegion} and \ref{tab:RLbyRegion_compare}, respectively.

One discovery of note that makes the implementation of the block kriging model in R fast and simple is the option to input a \emph{SpatialPolygonsDataFrame} as the ``newdata'' argument in the kriging functions of the {\tt gstat} package \citep{pebesma_multivariable_2004}. In particular, we can input the spatial polygons for our three hydrologic regions, and get the averaged estimates directly. When polygons are input as the new data to krige, the function uses {\tt sp::spsample} to randomly sample points uniformly across each polygon and calculates a block average \citep{pebesma_multivariable_2004, bivand_applied_2013}. This is intuitively the same calculation we are doing, without specifying a grid of points beforehand.

This built-in function implementation runs much faster than kriging to a fine grid and then averaging over the points within each region. For example, when cokriging the shape and log(scale) parameters, kriging to a fine grid of 19,791 points and then averaging the gridded points by region took over 15 minutes while kriging to the regions took only 3.7 seconds, a substantial improvement to computation time. One could definitely use a coarser grid to speed things up, but the {\tt gstat} functions ({\tt gstat::predict} and {\tt gstat::krige} for multivariate and univariate kriging, respectively) with ``newdata'' set to the region polygons performs quickly and produces very similar estimates to the gridded version with 19,791 points. We compare the averaged parameter estimates for each region in Supplemental Table 1, where the values obtained from using the {\tt gstat} function shortcut are displayed in brackets. Due to the similarity of the estimates (with the largest deviation being 0.16 in the scale parameter), we evaluate all future results using the faster method.

Additionally, since the PARE and block kriging models both perform modeling on the log of the scale parameter, the estimates from these models must be translated back to the regular scale. Since these estimates are derived from the idea of taking the mean of the log(scale), we use approximations to bring it back to the mean of the scale as opposed to just taking the exponential value of the estimate. We use Taylor series approximations to transform back to the scale, which are outlined and derived in the appendix of \cite{fagnant_spatiotemporal_2021}.

After fitting each of our proposed models to the last 40 years of data, we compare results for the estimates across the three regions by looking at both the extreme value parameter estimates as well as the estimated return levels. In Table \ref{tab:paramByRegion}, we see that the scale parameter estimates for the PARE model and block kriging are similar, whereas those of the regional max model are much larger, by roughly 50\%. For all three models, the shape parameter for region three is the smallest. Similarly, region three usually has the largest scale parameter. We also note that the rate parameter estimates vary the least across regions or models, which provides the justification for holding the rate parameter constant in the simulations in Section \ref{sec:sim_results}. The approximately constant estimates suggest the modeling of this parameter might be weighed against the added complexity of including it. If we want to simplify modeling, we could hold this parameter constant across regions.

% rate (aka lam) does come into the calculation of the confidence interval of the return level, but does not require uncertainty measures (var-cov)

From the estimated extreme value parameter estimates, we also calculate the 25-, 100-, and 500-year return level estimates (equivalent to the 4\%, 1\%, and 0.2\% precipitation events, respectively) and display the results in the first 9 rows of Table \ref{tab:RLbyRegion_compare}. The regional max model produces the largest return level estimates, but has somewhat similar results to the PARE model. For the PARE model, region one exhibits the largest return levels. The return levels for region three in the block kriging approach are lower than expected. 

An important distinction in the return level estimates is that the PARE model produces smaller standard errors, illustrating the viability of our approach. However, in order to calculate standard errors and confidence intervals for the return levels, one piece of information we need is the covariance matrix of the shape and scale parameters. Due to the nature of modeling the parameters separately in the PARE model, we set the covariance to be zero, despite the parameters being negatively correlated. Future development of this model will address joint models the two parameter estimates, so that the covariance can be used to improve our knowledge of the precision of the estimates.

% from this, also calculate return levels/ average exceedance probabilities!

We extend our modeling of return levels for each region to the temporal setting by incorporating the moving window approach used in \cite{fagnant_characterizing_2020}. Specifically, we repeat our modeling on data subset to different 40-year periods. In order to get a general idea of how these regional return level estimates have changed over time, we choose three 40-year windows spaced evenly across our period of record. The overlapping windows chosen are 1921-1960, 1951-1990, and 1981-2020. We provide tables of the parameter estimates for the last window in Table \ref{tab:paramByRegion}, and the first and second windows in Supplemental Tables . 

\begin{figure}
    \centering
    \includegraphics[width=.9\textwidth]{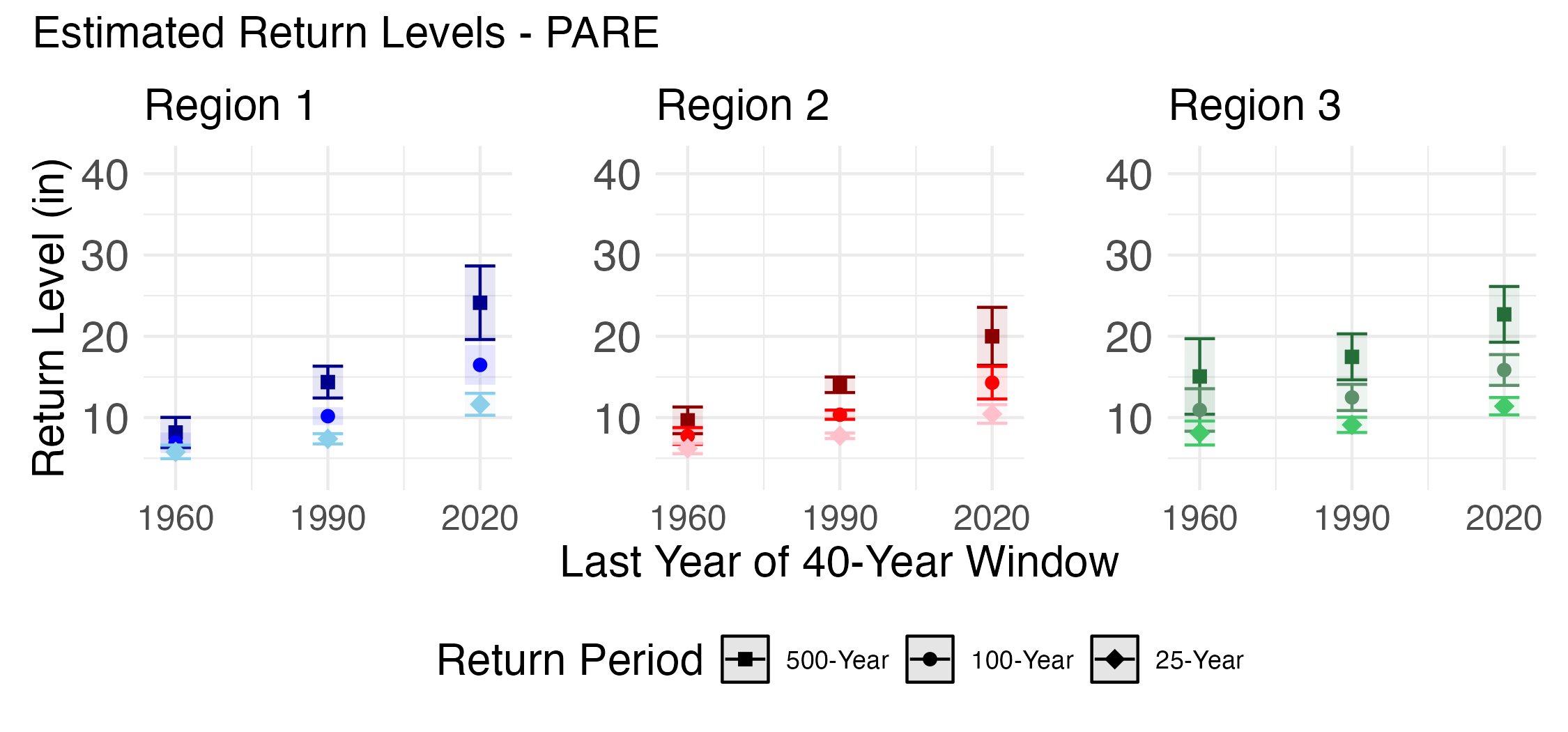}
    \caption{Return level estimates and 95\% CIs for moving windows - PARE Model. Regions are identified in Figure \ref{fig:Stations_map}, and color coded in the same fashion.}
    \label{fig:Model1_windows}
\end{figure}

\begin{figure}
    \centering
    \includegraphics[width=.9\textwidth]{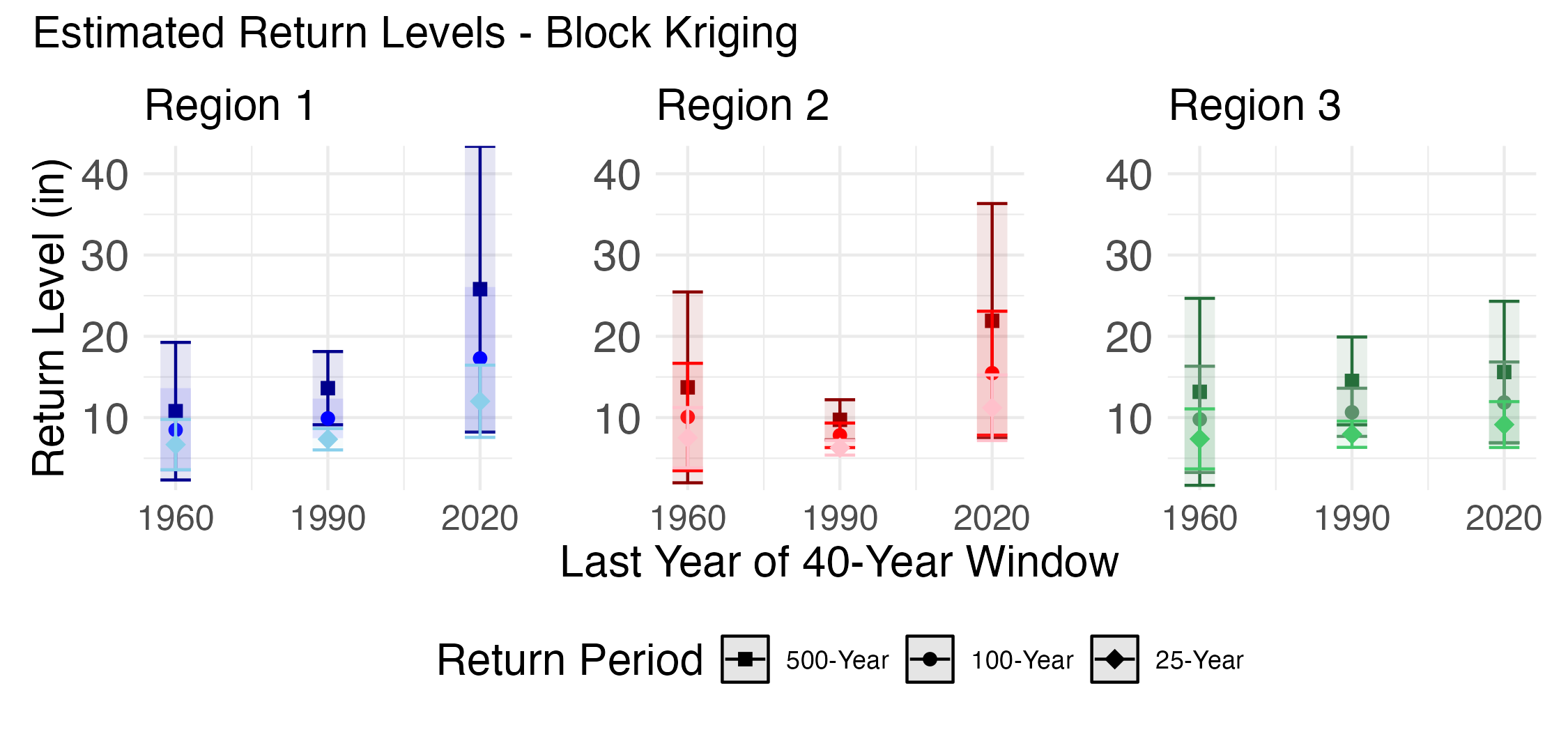}
    \caption{24-hour Return level estimates and 95\% CIs for moving windows - Block Kriging. Regions are identified in Figure \ref{fig:Stations_map}, and color coded in the same fashion. Values for visualization extracted from \cite{fagnant_spatiotemporal_2021}}
    \label{fig:Model2_windows}
\end{figure}

\begin{figure}
    \centering
    \includegraphics[width=.9\textwidth]{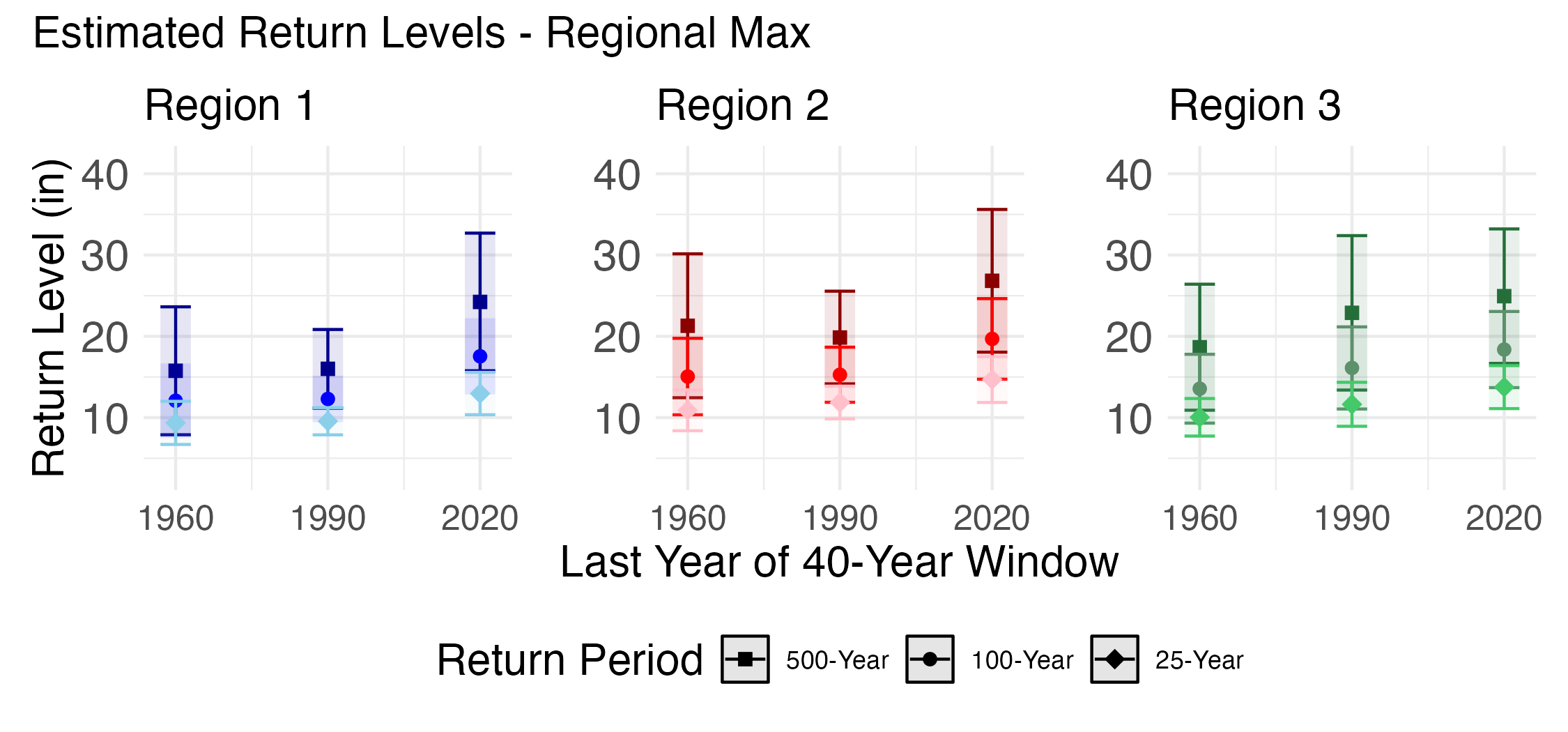}
    \caption{Return level estimates and 95\% CIs for moving windows - Regional Max. Regions are identified in Figure \ref{fig:Stations_map}, and color coded in the same fashion. Values for visualization extracted from \cite{fagnant_spatiotemporal_2021}}
    \label{fig:Model3_windows}
\end{figure}

We display return level estimates along with standard errors for the three models for each window of time. See Supplemental Table 4 for 1921-1960 and Supplemental Table 5 for 1951-1990.  To visualize clearly how return levels have changed over time for the three hydrologic regions, we also plot the return level estimates by window as a simple time series, along with 95\% pointwise confidence intervals. Time series plots of the 25-, 100-, and 500-year return levels estimates with 95\% pointwise confidence intervals, for each of the three regions, are provided in Figures \ref{fig:Model1_windows} -- \ref{fig:Model3_windows}. 

Region one has similar return levels for the final period (ending in 2020) across all models. The return level is around 24-25 inches for the 500-year and around 17 inches for the 100-year. The uncertainty associated with the block kriging and regional max models, makes interpretation of trends more difficult. For example, region two saw slight decreases in return levels from 1960 to 1990 for the block kriging model and the regional max model, but the uncertainty is large. 

Considering the PARE model results, regions one and two exhibit a steady increase in return levels since 1921. The increase in return levels from 1990 to 2020 was greater than that from 1960 to 1990. Region one represents the area to the northwest of Houston, TX, and region two covers the center part of the city. Region three saw the most gradual increase in extreme rainfall over time. It is worth noting, that region three is coastal and includes Galveston Island, which is regularly inundated with heavy rains. 

% Model 1 - PARE Model

% Model 2 - Block Kriging

% Model 3 - Regional Max

Two interesting results are that the return level estimates for the PARE model demonstrate consistent trends, comparable to the findings in \cite{fagnant_characterizing_2020}, and also provide precise confidence intervals for the 25-, 100- and 500-estimated return levels for each region. For all regions and return levels, the PARE model has lower uncertainty estimates than the regional max and the block kriging approach. The regional max model in has very wide confidence intervals across all windows and regions. Similar to how we suspected the construction of the data using the maximum increases the return level estimates for the regional max model, it also increases the variability of these estimates and therefore the confidence interval widths.

There has been an effort in recent years to update extreme rainfall data in the Houston area. In particular, NOAA Atlas 14 released updated rainfall frequency estimates for all of Texas in 2018 \citep{perica_precipitation-frequency_2018}. This updated data set provides gridded precipitation frequency estimates (equivalent to return levels) across Texas for multiple durations (ranging from 5 minutes to 60 days) and multiple return periods (from 1 to 1000 years). We wish to compare our regional return level estimates to updated region estimates for the Harris County hydrologic regions using this updated NOAA Atlas 14 data. As of the time of writing, the only such study found which also aims to bring return level estimates to the region level is a report produced by the American Council of Engineering Companies (ACEC) in Houston in 2019 \citep{acechouston_recommendation_2019}. The goals of this study slightly differed from ours, as they used the NOAA Atlas 14 frequency estimates directly, and also use these estimates to determine ideal regions for Harris County (i.e. grouping the watersheds into regions themselves). This resulted in almost the same designation as that of the Harris County Flood Control District (HCFCD) manual \citep{storey_harris_2009} which we use, except that ACEC placed the Sims Bayou watershed in region 2 as opposed to region 3.

We compare our estimated regional return levels for the last 40-year window (1981-2020) to the ACEC study estimates in the last three rows of Table \ref{tab:RLbyRegion_compare}. We note that since regions 2 and 3 were slightly different between our analysis and theirs, we cannot compare these values directly, and so mark them with an asterisk. However, region 1 is designated exactly the same for both studies. We discover that our proposed PARE model matches the ACEC estimates more closely than the other models, indicating that the PARE model run on raw rainfall data performs similarly to the most recent advanced modeling of rainfall frequency estimates by NOAA. It is promising that estimates from our PARE model are consistent with these estimates derived from NOAA Atlas 14, because the NOAA report is the official source of precipitation frequency estimates provided for the U.S. government \citep{perica_precipitation-frequency_2018} and is widely used and accepted.

\section{Discussion}\label{sec:disc}
We brought forward an end-to-end approach to estimating the return level, focusing on 25-, 100-, and 500-year levels for Houston, Texas. Our approach made use of hierarchical spatial modeling, and introduced point-to-area-random-effects or PARE to address the change of support. Our multi-tiered approach provides return level estimates for key regions in flood modeling along with uncertainty quantification. The results are consistent over our three time epochs, representing a span of 100 years. When compared with the block kriging and regional max models, we obtain a clear picture of the rainfall dynamics in Houston, Texas. A key advantage of the PARE approach is the use of standard software, ensuring ease of use for future problems. The method also easily incorporates additional fixed effects. 

\section*{Supplementary Materials}
A pdf file containing the Supplemental Tables is included in the Supplemental Materials. Code to reproduce the PARE analysis for the three windows considered in this paper is provided as a Quarto project available for download on Github (\href{https://github.com/juliaSchedler/ST_PARE}{https://github.com/juliaSchedler/ST\_PARE}), or viewing as a rendered vignette at \href{https://juliaschedler.github.io/ST_PARE/}{https://juliaschedler.github.io/ST\_PARE/}. Due to large file sizes, additional resources on the GPD fitting process for all 80 years' worth of rolling windows and models used for comparison to PARE are provided in another Github Repository (\href{https://github.com/carly-fagnant/Spatial_Extreme_Value_Modeling}{https://github.com/carly-fagnant/Spatial\_Extreme\_Value\_Modeling}).

\bibliography{ST_PARE}

% Revision Table 1
% UPDATE of  what was previously TABLE 9

% TABLE XXXX
% Simulation results
% TABLE 9
\begin{table}
\resizebox{\textwidth}{!}{
\begin{tabular}{ccc@{\extracolsep{0.8cm}}c@{}c@{}c@{\extracolsep{0.8cm}}c@{}c@{}c@{\extracolsep{0.8cm}}c@{}c@{}c}

%{ccc@{\extracolsep{0.8cm}}c@{\extracolsep{12pt}}cc@{\extracolsep{0.8cm}}c@{\extracolsep{12pt}}cc}
% \cline{4-9}
\cline{3-5} \cline{6-8} \cline{9-11}&       & \multicolumn{3}{c}{Region 1} & \multicolumn{3}{c}{Region 2} & \multicolumn{3}{c}{Region 3} \\
% \cline{3} 
\cline{3-5} \cline{6-8} \cline{9-11}
&          & Mean     & RMSE     & MAE   & Mean     & RMSE     & MAE  & Mean     & RMSE     & MAE  \\ \hline
Scale
%       Truth     Mean     RMSE       MAE     Mean     RMSE       MAE     Mean     RMSE      MAE
% Reg1 233.64 233.9610 1.346618 1.0536345 235.8534 2.519408 2.2470368 250.8350 17.59930 17.19496
% Reg2 246.78 247.0729 1.312451 1.0536885 245.1087 2.128176 1.8336914 264.2653 17.83535 17.48534
% Reg3 229.38 229.7001 1.199355 0.9159539 229.6051 1.169656 0.9225165 247.0109 17.96241 17.63093
    %   Truth     Mean        RMSE       MAE     Mean       RMSE       MAE
& Truth & 233.64 &  &  & 246.78 &   & & 229.38& &  & \\
& PARE & 233.96 & \bf{1.3466} & \bf{1.0536} & 247.07 &  \bf{1.3125} & \bf{1.0537}& 229.70&1.1994 & \bf{0.9160} & \\
& Block Kriging & 235.85 & 2.5194 & 2.2470 & 245.11&2.1282 & 1.8337 & 229.61 & \bf{1.1697} & 0.9225 & \\
& Regional Max & 250.84 & 17.5993 & 17.1950 &264.27 & 17.8354 & 17.4853 & 247.01 & 17.9624 & 17.6309 & \\
                            \hline
Shape
%       Truth      Mean        RMSE         MAE      Mean        RMSE        MAE      Mean      RMSE       MAE
% Reg1 0.2044 0.2012761 0.005563777 0.004412464 0.1983770 0.007242354 0.00646080 0.3326313 0.1291051 0.1282313
% Reg2 0.2319 0.2295365 0.005029726 0.004081760 0.2260699 0.007253734 0.00612616 0.3675515 0.1363450 0.1356515
% Reg3 0.1641 0.1620822 0.004367115 0.003496306 0.1699796 0.007085289 0.00603868 0.2774722 0.1142810 0.1133722
    %   Truth      Mean         RMSE    MAE      Mean       RMSE      MAE
& Truth & 0.2044 &  &  & 0.2319 &   & & 0.1641 & &  & \\
& PARE & 0.2013 & \bf{0.0056} & \bf{0.0044} & 0.2295 & \bf{0.0050} & \bf{0.0041} &  0.1621 & \bf{0.0044}& \bf{0.0035}& \\
& Block Kriging & 0.1984 & 0.0072 & 0.0065 & 0.2261 & 0.0073 & 0.0061 & 0.1700 & 0.0071 & 0.0060 & \\
& Regional Max & 0.3326 & 0.1291 & 0.1282 & 0.3676 & 0.1363 & 0.1357 & 0.2775 & 0.1143 & 0.1134 & \\
                            \hline
\end{tabular}}
\caption{Extreme value parameter estimates from our proposed models tested on the simulated data for 50 iterations. Bolding denotes the smallest value of either RMSE or MAE for estimating the given parameter for a given region. Note that the rate parameter is held constant, so rate estimates are not evaluated here. Included are the true parameter values used for simulation (Truth), as well as the mean parameter estimates (Mean), root mean squared error (RMSE), and mean absolute error (MAE) for each region and each model from the 50 iterations. Smaller values of RMSE and MAE indicate better performance of a model at recovering the true parameter values. Our proposed PARE Model performs almost uniformly better according to these metrics.}
% \caption{Simulation results for Models 1 and 2, using 50 iterations.}
\label{tab:simTable}
\end{table}

% TABLE 2
\begin{table}
\centering
\begin{tabular}{ccccc}
\toprule    %\hline
Method                   & Parameter & Region 1 & Region 2 & Region 3 \\ 
\toprule  %\hline
% Model 1
1. PARE Model  
&scale&215.08(4.3232)&236.68(4.8949)&223.99(3.6994)\\
&shape&0.2(0.0164)&0.16(0.0168)&0.19(0.0134)\\
&rate&0.06(0.0023)&0.04(0.0023)&0.06(0.0019)\\
                         \hline
% Model 2
2. Block Kriging 
%               Reg1         Reg2         Reg3
% scale 208.83406078 228.29669840 229.40551739
%    SE        Reg1        Reg2        Reg3
% scale 16.52064811 18.55700411 17.52987563
% shape  0.06956766  0.07149129  0.06718452
% rate  0.005772753 0.005789695 0.005746261
& scale     & 208.83 (16.521) & 228.30 (18.557) & 229.41 (17.530) \\
& shape     & 0.2205 (0.0696) & 0.1771 (0.0715) & 0.1121 (0.0672) \\
& rate      & 0.0559 (0.0058) & 0.0554 (0.0058) & 0.0557 (0.0057) \\ 
                         \hline
% Model 3 
3. Regional Max 
% consol_fit
%               Reg1         Reg2         Reg3
% scale 291.20737755 343.46122206 331.37685058
% shape   0.15229146   0.13974427   0.13513157
% rate    0.05653662   0.06009582   0.05571526
%   SE         Reg1        Reg2        Reg3
% scale 14.89786046 16.77782654 16.74679031
% shape  0.03768466  0.03546599  0.03664504
% > sqrt(fitreg1$rate*(1-fitreg1$rate)/14610)
%  0.001910743  0.00196625  0.001897638
& scale     & 291.21 (14.897) & 343.46 (16.778) & 331.38 (16.747) \\
& shape     & 0.1523 (0.0377) & 0.1397 (0.0355) & 0.1351 (0.0366) \\
& rate      & 0.0565 (0.0019) & 0.0601 (0.0020) & 0.0557 (0.0019) \\ % rate has no SE for this model - NOT TRUE
\bottomrule         %\hline
\end{tabular}
\caption{Comparison of extreme value parameter estimates (standard errors in parentheses) across our proposed models for the last 40 years of data, 1981-2020. As we predicted, the regional max results differ more than the other two models with its scale parameters being much higher, which may lead to larger return level estimates.}
\label{tab:paramByRegion}
\end{table}

% TABLE 8
% Return levels - window 3- with model names, ACEC Standards
\begin{table}
\centering
% \resizebox{\textwidth}{!}{
\begin{tabular}{ccccc}
\toprule    %\hline
Method                  & Return Period & Region 1 & Region 2 & Region 3 \\ 
\toprule  %\hline
% Model 1
1. PARE Model 
&25-Year&11.634(0.6874)&10.442(0.5857)&11.399(0.5461)\\
&100-Year&16.48(1.2435)&14.291(1.0239)&15.856(0.9666)\\
&500-Year&24.131(2.3097)&20.002(1.8186)&22.7(1.7504)\\
                         \hline
% Model 2
2. Block Kriging
%       25-yr RL       SE
% Reg1 12.008052 2.266245
% Reg2 11.209502 2.057745
% Reg3  9.141326 1.442890
%      100-yr RL       SE
% Reg1  17.27747 4.484813
% Reg2  15.46461 3.890739
% Reg3  11.86537 2.544536
%      500-yr RL       SE
% Reg1  25.80012 8.969202
% Reg2  21.91129 7.363977
% Reg3  15.60741 4.437395
& 25-Year   & 12.01 (2.27) & 11.21 (2.06) &  9.14 (1.44) \\
& 100-Year  & 17.28 (4.48) & 15.46 (3.89) & 11.87 (2.54) \\
& 500-Year  & 25.80 (8.97) & 21.91 (7.36) & 15.61 (4.44) \\ 
                         \hline
% Model 3 
3. Regional Max 
%      25-yr RL       SE
% Reg1 12.95903 1.331772
% Reg2 14.68239 1.435428
% Reg3 13.75276 1.353835
%      100-yr RL       SE
% Reg1  17.54066 2.391334
% Reg2  19.67649 2.531448
% Reg3  18.37010 2.387069
%      500-yr RL       SE
% Reg1  24.22698 4.318485
% Reg2  26.82833 4.484441
% Reg3  24.93663 4.220351
& 25-Year  & 12.96 (1.33) & 14.68 (1.44) & 13.75 (1.35) \\
& 100-Year & 17.54 (2.39) & 19.68 (2.53) & 18.37 (2.39) \\
& 500-Year & 24.23 (4.32) & 26.83 (4.48) & 24.94 (4.22) \\
        \hline \hline
ACEC Estimates
& 25-Year  & 10.90 & 11.50* & 12.30* \\
& 100-Year & 16.30 & 16.90* & 18.00* \\
& 500-Year & 24.20 & 25.00* & 27.20* \\
\bottomrule 
\end{tabular}%}
\caption{Comparison of return level estimates (in inches) across the proposed models for the last 40 years of data, 1981-2020. Standard error estimates are displayed in parentheses. Additionally, the model estimates are compared to region estimates produced by the American Council of Engineering Companies (ACEC) in Houston \citep{acechouston_recommendation_2019}. The ACEC estimates for regions 2 and 3 are marked with asterisks because the hydrologic regions are slightly different than the HCFCD designation. Our proposed PARE Model produces very similar estimates to the ACEC study for region 1, which is the same as our region 1.}
\label{tab:RLbyRegion_compare}
\end{table}

\end{document}

\title{Supplemental Tables for \emph{Spatial-Temporal Extreme Modeling for Point-to-Area Random Effects (PARE)}}
\maketitle
% TABLE S1
\begin{table}[h]
\centering
\begin{tabular}{@{}cccccc@{}}
\toprule
Parameter  & Region 1    & Region 2    & Region 3                \\ \midrule
scale  &  208.34 [208.18] & 227.62 [227.54] & 228.68 [228.73] \\
shape  &  0.2199 [0.2205] & 0.1776 [0.1771] & 0.1109 [0.1121] \\
rate   &  0.0559 [0.0559] & 0.0554 [0.0554] & 0.0557 [0.0557] \\
\bottomrule
\end{tabular}
\caption{Comparison of parameter estimates from kriging to a fine grid and averaging the gridded values versus using the shortcut in the {\tt gstat} package to krige to the polygons of our regions. The latter method's estimates are in brackets. Note that the estimates are very close, so faster computation times lead us to use the built-in functionality of {\tt gstat}.}
\label{tab:compareAvgMethod}
\end{table}

% TABLE S2
% Return levels, window 3, 1981-2020
\begin{table}
\centering
\begin{tabular}{ccccc}
\toprule    %\hline
Method                   & Return Period & Region 1 & Region 2 & Region 3 \\ 
\toprule  %\hline
\multirow{1. PARE Model} 
&25-Year&11.634(0.6874)&10.442(0.5857)&11.399(0.5461)\\
&100-Year&16.48(1.2435)&14.291(1.0239)&15.856(0.9666)\\
&500-Year&24.131(2.3097)&20.002(1.8186)&22.7(1.7504)\\
                         \hline
\multirow{2. Block Kriging}
& 25-Year   & 12.008 (2.266) & 11.210 (2.058) &  9.141 (1.443) \\
& 100-Year  & 17.277 (4.485) & 15.465 (3.891) & 11.865 (2.545) \\
& 500-Year  & 25.800 (8.969) & 21.911 (7.364) & 15.607 (4.437) \\ 
                         \hline
\multirow{3. Regional Max} 
%      25-yr RL       SE
& 25-Year  & 12.959 (1.332) & 14.682 (1.435) & 13.753 (1.354) \\
& 100-Year & 17.541 (2.391) & 19.676 (2.531) & 18.370 (2.387) \\
& 500-Year & 24.227 (4.318) & 26.828 (4.484) & 24.937 (4.220) \\
\bottomrule       
\end{tabular}
\caption{Comparison of return level estimates (in inches) across our proposed models for the last 40 years of data, 1981-2020. Standard error estimates are displayed in parentheses.}
\label{tab:RLbyRegion}
\end{table}

% TABLE S3
% Parameters - Window 1-- 1921-1960
\begin{table}
\centering
\begin{tabular}{ccccc}
\toprule    %\hline
Method                   & Parameter & Region 1 & Region 2 & Region 3 \\ 
\toprule  %\hline
\multirow{1. PARE Model} 
%               Reg1         Reg2         Reg3
% scale 229.39636929 199.97938701 199.06224578
%   SE       Reg1       Reg2       Reg3
% scale 8.51002107 3.99609682 9.26698627
% shape 0.05080284 0.02671465 0.06465595
% rate 0.00070848  0.00038276  0.00088771
&scale&230.86(5.3213)&200.16(3.0221)&198.46(4.6)\\
&shape&-0.02(0.0314)&0.05(0.0205)&0.15(0.0316)\\
&rate&0.03(6e-04)&0.03(4e-04)&0.03(6e-04)\\
                         \hline
\multirow{2. Block Kriging} 
%               Reg1         Reg2         Reg3
% scale 208.15867142 194.78245904 196.99111770
%    SE        Reg1        Reg2        Reg3
% scale 12.24796821 10.79963398 12.50849713
% shape  0.09460119  0.09251395  0.09758864
& scale & 208.16 (12.248) & 194.78 (10.800) & 196.99 (12.508) \\
& shape & 0.0720 (0.0946) & 0.1382 (0.0925) & 0.1274 (0.0976) \\
& rate  & 0.0304 (0.0010) & 0.0304 (0.0010) & 0.0304 (0.0010) \\ 
                         \hline
\multirow{3. Regional Max} 
%               Reg1         Reg2         Reg3
% scale 247.98127025 240.97976262 237.29743389
% shape   0.10208539   0.17219320   0.14915125
% rate    0.05130079   0.04298426   0.04374078
%   SE        Reg1        Reg2       Reg3
% scale 19.96427696 14.56145681 14.6899729
% shape  0.06088692  0.04584126  0.0477731
% > sqrt(fitreg1_22$rate*(1-fitreg1_22$rate)/fitreg1_22$n)
%  0.002667072  0.001677989  0.001713683
& scale & 247.98 (19.964) & 240.98 (14.561) & 237.30 (14.690) \\
& shape & 0.1021 (0.0609) & 0.1722 (0.0458) & 0.1492 (0.0478) \\
& rate  & 0.0513 (0.0027) & 0.0430 (0.0017) & 0.0437 (0.0017) \\ 
\bottomrule         %\hline
\end{tabular}
\caption{Extreme value parameter estimates (standard errors in parentheses) from our proposed models for the first 40-year window, 1921-1960.}
\label{tab:paramByRegion_window22}
\end{table}

% TABLE S4
% Parameters - Window 2, 1961-1980
\begin{table}
\centering
\begin{tabular}{ccccc}
\toprule    %\hline
Method                   & Parameter & Region 1 & Region 2 & Region 3 \\ 
\toprule  %\hline
\multirow{1. PARE Model} 
%               Reg1         Reg2         Reg3
% scale 173.61625117 206.61060644 213.62969470
%    SE       Reg1       Reg2       Reg3
% scale 5.48393386 3.24363632 7.82049610
% shape 0.02772861 0.01362721 0.03226143
% rate  0.00062725 0.00031301 0.00072602
&scale&171.96(4.4827)&202.32(2.7176)&217.8(6.6254)\\
&shape&0.17(0.0131)&0.13(0.0068)&0.16(0.0153)\\
&rate&0.03(7e-04)&0.03(3e-04)&0.03(8e-04)\\
                         \hline
\multirow{2. Block Kriging} 
%               Reg1         Reg2         Reg3
% scale 182.42895977 205.61173247 203.37340679
%    SE       Reg1       Reg2        Reg3
% scale 8.44916373 8.64449874 10.73431585
% shape 0.03949913 0.03579795  0.04486861
& scale & 182.43 (8.449) & 205.61 (8.644) & 203.37 (10.734) \\
& shape & 0.1491 (0.0395) & 0.0485 (0.0358) & 0.1388 (0.0449) \\
& rate  & 0.0306 (0.0009) & 0.0320 (0.0009) & 0.0327 (0.0010) \\ 
                         \hline
\multirow{3. Regional Max} 
%               Reg1         Reg2         Reg3
% scale 249.41072546 318.74883578 243.08664829
% shape   0.10071960   0.09638578   0.17749802
% rate    0.05701574   0.05927447   0.05078713
%    SE        Reg1        Reg2        Reg3
% scale 12.47149773 15.26984070 13.87518803
% shape  0.03612175  0.03377189  0.04437016
% > sqrt(fitreg1_52$rate*(1-fitreg1_52$rate)/fitreg1_52$n)
% 0.001918335  0.00195362  0.001816492
& scale & 249.41 (12.471) & 318.75 (15.270) & 243.09 (13.875) \\
& shape & 0.1007 (0.0361) & 0.0964 (0.0338) & 0.1775 (0.0444) \\
& rate  & 0.0570 (0.0019) & 0.0593 (0.0020) & 0.0508 (0.0018) \\ 
\bottomrule         %\hline
\end{tabular}
\caption{Extreme value parameter estimates (standard errors in parentheses) from our proposed models for the second 40-year window, 1951-1990.}
\label{tab:paramByRegion_window52}
\end{table}

% TABLE S5
% Return Levels Window 1- 1921-1960
\begin{table}
\centering
% \resizebox{\textwidth}{!}{
\begin{tabular}{ccccc}
\toprule    %\hline
Method                   & Return Period & Region 1 & Region 2 & Region 3 \\ 
\toprule  %\hline
\multirow{1. PARE Model}
%      25-yr RL        SE
% Reg1 4.636148 0.4814777
% Reg2 4.999941 0.3049168
% Reg3 8.118740 1.5453151
%      100-yr RL        SE
% Reg1  5.229928 0.6675198
% Reg2  5.864730 0.4499356
% Reg3 10.962001 2.7366548
%      500-yr RL        SE
% Reg1  5.805805 0.8922391
% Reg2  6.814874 0.6475776
% Reg3 15.091936 4.8688996
&25-Year&5.781(0.4269)&6.2(0.3274)&8.102(0.7539)\\
&100-Year&6.9(0.645)&7.733(0.5268)&10.938(1.3344)\\
&500-Year&8.161(0.9508)&9.66(0.8373)&15.059(2.3736)\\
                            \hline
\multirow{2. Block Kriging} 
%      25-yr RL       SE
% Reg1 6.681097 1.578415
% Reg2 7.523245 1.895190
% Reg3 7.375858 1.885067
%      100-yr RL       SE
% Reg1  8.472337 2.629995
% Reg2 10.073333 3.372384
% Reg3  9.784245 3.334681
%      500-yr RL       SE
% Reg1  10.78897 4.319249
% Reg2  13.71672 5.990183
% Reg3  13.16973 5.869584
& 25-Year   &  6.681 (1.578) &  7.523 (1.895) &  7.376 (1.885) \\
& 100-Year  &  8.472 (2.630) & 10.073 (3.372) &  9.784 (3.335) \\
& 500-Year  & 10.789 (4.319) & 13.717 (5.990) & 13.170 (5.870) \\ 
                            \hline
\multirow{3. Regional Max} 
%       25-yr RL       SE
% Reg1  9.349091 1.354207
% Reg2 10.895254 1.283326
% Reg3 10.037471 1.179219 
%      100-yr RL       SE
% Reg1  12.07285 2.340859
% Reg2  15.04967 2.401693
% Reg3  13.55301 2.161059
%      500-yr RL       SE
% Reg1  15.75856 4.015606
% Reg2  21.29709 4.516533
% Reg3  18.65929 3.956183
& 25-Year  & 9.349 (1.354) & 10.895 (1.283) & 10.037 (1.179) \\
& 100-Year & 12.073 (2.341) & 15.050 (2.402) & 13.553 (2.161) \\
& 500-Year & 15.759 (4.016) & 21.297 (4.517) & 18.659 (3.956) \\
\bottomrule         %\hline
\end{tabular}%}
\caption{Return level estimates (in inches) from our proposed models for the first 40-year window, 1921-1960. Standard error estimates are displayed in parentheses.}
\label{tab:RLbyRegion_window22}
\end{table}

% TABLE S6
% Return Levels Window 2 1961-1980
\begin{table}
\centering
\begin{tabular}{ccccc}
\toprule    %\hline
Method                   & Return Period & Region 1 & Region 2 & Region 3 \\ 
\toprule  %\hline
\multirow{1. PARE Model} 
%      25-yr RL        SE
% Reg1 7.333369 0.6023992
% Reg2 6.379354 0.2357537
% Reg3 9.378770 0.9576031
%      100-yr RL        SE
% Reg1 10.047392 1.0824592
% Reg2  7.962832 0.3731154
% Reg3 13.022058 1.7249567
%      500-yr RL        SE
% Reg1 14.095327 1.9670303
% Reg2  9.954234 0.5869751
% Reg3 18.535749 3.1534281

&25-Year&7.391(0.3201)&7.759(0.1733)&9.104(0.4793)\\
&100-Year&10.177(0.5587)&10.354(0.2883)&12.477(0.8223)\\
&500-Year&14.367(0.9991)&14.038(0.4891)&17.472(1.4436)\\
                                        \hline
\multirow{2. Block Kriging} 
%      25-yr RL        SE
% Reg1 7.337291 0.6702666
% Reg2 6.288706 0.4619951
% Reg3 7.950432 0.8230699
%      100-yr RL        SE
% Reg1  9.899646 1.2439132
% Reg2  7.818000 0.7706997
% Reg3 10.650220 1.5113501
%      500-yr RL       SE
% Reg1 13.621299 2.299251
% Reg2  9.727501 1.259705
% Reg3 14.511105 2.757498
& 25-Year   &  7.337 (0.670) & 6.289 (0.462) &  7.950 (0.823) \\
& 100-Year  &  9.900 (1.244) & 7.818 (0.771) & 10.650 (1.511) \\
& 500-Year  & 13.621 (2.299) & 9.728 (1.260) & 14.511 (2.757) \\ 
                                        \hline
\multirow{3. Regional Max} 
%       25-yr RL        SE
% Reg1  9.552109 0.8540718
% Reg2 11.857890 1.0252344
% Reg3 11.636494 1.3806254
%      100-yr RL       SE
% Reg1  12.29505 1.455358
% Reg2  15.27193 1.729837
% Reg3  16.10920 2.577707
%      500-yr RL       SE
% Reg1  15.99911 2.469884
% Reg2  19.85230 2.909360
% Reg3  22.88928 4.847009
& 25-Year  & 9.552 (0.854) & 11.858 (1.025) & 11.636 (1.381) \\
& 100-Year & 12.295 (1.455) & 15.272 (1.730) & 16.109 (2.578) \\
& 500-Year & 15.999 (2.470) & 19.852 (2.909) & 22.889 (4.847) \\
\bottomrule         %\hline
\end{tabular}
\caption{Return level estimates (in inches) from our proposed models for the second 40-year window, 1951-1990. Standard error estimates are displayed in parentheses.}
\label{tab:RLbyRegion_window52}
\end{table}